# The influence of non-idealities on the thermoelectric power factor of nanostructured superlattices

Mischa Thesberg[1], Mahdi Pourfath[1], Hans Kosina[1], and Neophytos Neophytou[2*]

[1]Institute for Microelectronics, Technical University of Vienna,

Gußhausstraße 27-29 / E360, A-1040 Wien, Austria

[2]School of Engineering, University of Warwick, Coventry, CV4 7AL, UK

[*]N.Neophytou@warwick.ac.uk

## Abstract

Cross-plane superlattices composed of nanoscale layers of alternating potential wells and barriers have attracted great attention for their potential to provide thermoelectric power factor improvements and higher *ZT* figure of merit. Previous theoretical works have shown that the presence of optimized potential barriers could provide improvements to the Seebeck coefficient through carrier energy filtering, which improves the power factor by up to 40%. However, experimental corroboration of this prediction has been extremely scant. In this work, we employ quantum mechanical electronic transport simulations to outline the detrimental effects of random variation, imperfections and non-optimal barrier shapes in a superlattice geometry on these predicted power factor improvements. Thus we aim to assess either the robustness or the fragility of these theoretical gains in the face of the types of variation one would find in real material systems. We show that these power factor improvements are relatively robust against: overly thick barriers, diffusion of barriers into the body of the wells, and random fluctuations in barrier spacing and width. However, notably, we discover that extremely thin barriers and random fluctuation in barrier heights by as little as 10% is sufficient to entirely destroy any power factor benefits of the optimized geometry. Our results could provide performance optimization routes for nanostructured thermoelectrics and elucidate the reasons why significant power factor improvements are not commonly realized in superlattices, despite theoretical predictions.

# I. Introduction

The thermoelectric performance of a material is quantified by the dimensionless figure of merit $ZT=\sigma S^2 T/\kappa$, where $\sigma$ is the electrical conductivity, $S$ is the Seebeck coefficient, and $\kappa$ is the thermal conductivity. Large improvements in the $ZT$ of nanostructures due to the reduction of the thermal conductivity have recently been demonstrated [1]. Similar benefits from power factor ($\sigma S^2$) improvements, however, have not yet been realized. This is attributed to the adverse interdependence of the electrical conductivity and Seebeck coefficient via the carrier density, which proves very difficult to overcome. To achieve power factor improvements, current efforts revolve around engineering the density of states of low-dimensional materials [2-6], modulation doping [7-10], introducing energy resonances in the density of states [11-12], and energy filtering in nanocomposites and superlattices [13-22]. Although theoretical works indicate that power factor improvements are possible, to-date experiments do not commonly demonstrate significant success in realizing these improvements. With respect to nanostructured superlattice structures specifically (one of the most promising and discussed methods), only improvements in the Seebeck coefficient and not the power factor have been experimentally observed [20].

In this work, we employ quantum mechanical electronic transport simulations to provide a critical examination of the potential of nanostructured cross-plane superlattices, to provide power factor improvements in the presence of non-idealities. Cross plane superlattices consist of alternating nanoscale material layers that form potential wells and barriers along the transport direction (see Fig. 1). Previous theoretical works by us and others have identified that such geometries can be optimized to achieve a larger power factor compared to a uniform material by up to 40% [23-25]. However, experimental verification of such power factor gains have not been forthcoming. In consideration of this scarcity of experimental corroboration, we attempt here to explore the effect of random fluctuations and variations from optimized geometry on these theoretical gains. Such imperfections are inevitable in any real system and we seek to assess the fragility or robustness of predicted power factor improvements in the face of such non-idealities.

Starting from such an optimized geometry, we examine the influence of a series of structure non-idealities on the power factor. Specifically, we consider the effect of imperfections of the barrier and well shapes (deviations from the square well/barrier shape, which as we show is the ideal shape), fluctuations in the well and barrier widths, and fluctuations in the barriers' heights. We show that statistical fluctuations of these parameters have the potential to entirely negate the power factor benefits that the ideal, optimal, superlattice geometry offers. Particularly detrimental to the power factor are: i) random fluctuations in the barrier heights, which can cause power factor reduction to values even below those of the uniform material, and ii) ultra-thin barriers, which allow significant quantum mechanical tunneling, thus eroding the Seebeck gains brought by the barriers.

## II. Approach

As both quantum tunnelling and the energy mixing effects of electron-phonon interactions are crucial considerations in energy-filtering systems [14], we use here the non-equilibrium Green's function (NEGF) approach, including the effect of electron scattering with acoustic and optical phonons [26-27]. The system is treated as 1D channel within the effective mass model. The effect of electron scattering with acoustic and optical phonons in NEGF is modeled by including a self-energy on the diagonal elements of the Hamiltonian. This approximation has been shown to be quantitatively valid for many systems [28], such as electrons in silicon [29], transport in carbon nanotubes [30], and many more, and captures all essential scattering physics. The convergence criteria for the ensuing self-consistent calculation was chosen to be current conservation. Thus, current is guaranteed to be conserved along the length of the channel to within 1% in the data shown here.

The strength of the electron-phonon coupling is given by $D_0$ as described in detail in [24-27]. This parameter, which has units of eV$^2$, represents the weighting of the Green's Function contributions to the scattering self-energy and is not to be confused with the phonon deformation potentials. The relationship between $D_0$ and the deformation potentials can be found in Ref. [30]. Since the purpose of this work is to illuminate the effect of non-ideal random imperfections in the potential barrier shapes and well shapes on power factor improvements, we do not consider other parameters that can vary in a real superlattice material, such as atomistic defects, strain fields, bandstructure changes in different regions of the potential wells for both electrons and phonons, etc. Thus, for the purposes of this work, we assume a constant effective mass throughout the material in all wells and barriers of value $m^*=m_0$, where $m_0$ is the rest mass of the electron, and a uniform phonon coupling constant $D_0$, which is taken to be the same for both acoustic and optical phonons for simplicity.

The power factor, $GS^2$, was obtained from the expression:

$$I = G\Delta V + SG\Delta T. \qquad (1)$$

For each value of the power factor, the calculation was run twice, initially with a small potential difference and no temperature difference ($\Delta T=0$), which yields the conductance ($G=I_{(\Delta T=0)}/\Delta V$), then again with a small temperature difference and no potential difference ($\Delta V=0$), which yields the Seebeck coefficient ($S=I_{(\Delta V=0)}/G\Delta T$). This method is validated in Ref. [24]. The requirement of current conservation throughout the system was the convergence criteria used to determine self-consistency of the scattering self-energy. A convergence value of 1% was chosen (i.e. convergence is reached if the current varies by no more than 1% along the length of the channel). As is common practice, only the imaginary part of the scattering self-energy included. The sharp features of the system required an unusually large number (~1000s) of convergence steps. For the data related to random variations, at least 100 different device structures were simulated overall. The exact number varies and reflects the amount needed to get relative convergence in the standard deviations (i.e. error bars) shown. The relevant matrix problems were solved using the recursive Green's function (RGF) method [31].

Figure 1 shows the superlattice band diagram under consideration. The Fermi level is denoted by the yellow-dashed line. The colormap shows the typical energy current spectrum. Most current flows over the potential barriers as expected, however, carrier energy relaxation due to the emission of optical phonons is observed within the potential wells (red thin line shows the average energy of the right going carriers). The slightly lower value of the average energy at the far edges of the channel is because the contacts are assumed to be a semi-infinite uniform bulk material without any barriers. Thus, the carrier energy will tend to relax to the band edge.

Channel calibration: In previous theoretical works [23-25,32] the optimal geometrical and material parameters for the highest thermoelectric power factor were identified as follows: i) the carrier energy within the potential wells needs to be semi-relaxed (i.e. the carriers only partially relax their energy in a potential well before reaching the next barrier), ii) the Fermi level needs to be placed high into the bands for improved conductivity, and ~$k_BT$ below the maximum of the barriers, iii) the width of the barriers needs to be large enough to prevent tunneling, but small enough to keep the channel resistance low. Following these design guidelines, we calibrate the superlattice material under consideration as follows (see Fig. 1): We set the well widths at $L_W$=20 nm, the barrier widths at $W$=3 nm, use perfect square shaped wells/barriers, place $E_F$=0.14 eV above the well conduction band which provides the highest ballistic conductance, place $V_B$=0.16 eV (~$k_BT$ above $E_F$) where $V_B$ is the height of the barriers. The value of $D_0$ is chosen such that the conductance of a 20 nm channel is found to be 50% of the ballistic value. This effectively amounts to fixing a mean free path of 20 nm for the system. The appropriate $D_0$ was found to be $D_0$=0.0016 eV$^2$ (which, again, is taken to be the same for acoustic and optical phonons).

The power factor and Seebeck coefficient are calculated using the method described by Kim *et* al. in Ref. [24]. The transport simulation is run twice, one with a small voltage difference Δ$V$ between the left and right terminals to determine the conductance, and then again with a small temperature difference Δ$T$, which, coupled with the conductance, is used to calculate the Seebeck coefficient and then the power factor.

## III. Results

Gaussian-shaped barrier channels: Once the channel is calibrated, we begin our first investigation of the influence of non-idealities with the simplest case of a Gaussian barrier (rather than a perfect square) with the single free parameter the variance which controls the barrier thickness. Such profiles can be formed when the doping profile is non-uniform in the channel, or the well-barrier interface is not sharp. This initial barrier shape, while simple, interpolates between two important limiting cases: that of a very thin Dirac δ-function like barrier and the case of a single solid barrier (see insets of Fig. 2c). We call this limiting case of a single solid barrier the 'bulk thermoelectric case'. The value in this extreme is represented in all appropriate figures with a magenta line. This line is important as when power factor values below this line are obtained, the superlattice structure approach has utterly failed and the material is in fact performing worse than the bulk thermoelectric material. The results for the thermoelectric coefficients (conductance $G$, Seebeck coefficient $S$, and thermoelectric power factor $GS^2$) versus the variance of the Gaussian profile are shown in Fig. 2. A very narrow δ-shaped barrier (small variance) will allow a significant degree of quantum mechanical tunneling, which will improve the channel conductance, but reduce the Seebeck coefficient [23]. Profiles with large variances reduce the conductance, but increase the Seebeck coefficient. Thus, a power factor of up to $GS^2 \sim 2.56 \times 10^{-14}$ W/K$^2$ can be achieved for moderate variance values (around Var~3.5 nm$^2$ in Fig. 2c). This is a similar value obtained in the case of the channel consisting of perfect square barriers/wells (the geometry shown in Fig. 1). Thus, we see that optimally chosen Gaussian parameters can produce moderate power factor gains (here on the order of ~20%) above a bulk thermoelectric material (shown by the magenta line in Fig. 2c). The benefit over the bulk thermoelectric case arises because the wells of the channel allow for high energy carriers with increased velocities, compared to low energy carriers in the single barrier geometry [13]. The wells locally increase the conductance, but reduce the Seebeck coefficient. Overall, however, the superlattice geometry provides a power factor advantage for the middle values of variance. Overly thin barriers, however, perform substantially worse than bulk thermoelectric materials. As the variance increases, on the other hand, the geometry starts to look like the single barrier geometry, and the power factors of the two geometries tend to converge. We expect this insight to be generally true regardless of specific material properties. In addition, in these results, we only use the optimal barrier height for high power factors. This behavior with respect to variance was found to be true even when the barrier height was changed (not shown) and thus it is also a general behaviour, independent of barrier height.

Curve-shaped well channels: Although the Gaussian case is very illuminating, one cannot separate tunnelling degradation due to thin barriers, from degradation due entirely due to transport in the well itself. In order to isolate the influence of the well shape independently of the effects of tunneling we examine a different geometry in which the shape and width of the potential barriers are square and fixed, but the shape of the well alone is now distorted as shown in the insets of Fig. 3c. Again, this could arise from non-uniform doping distribution in the wells or from diffusion of dopants from a superlattice, perhaps under the effects of annealing. We describe the shape of the wells by an exponential function $V_B \exp(-x/\xi)$, where $\xi$ is the decay length of the potential from the barrier top into the well. For $\xi=0$ nm we recover the perfect square well, and for large values we recover the uniform single barrier geometry, again represented by a magenta-dashed line. Figures 3a, 3b, and 3c show the dependence of the conductance, Seebeck coefficient and power factor on $\xi$, respectively. The black-dashed lines indicate the thermoelectric coefficients of the geometry with the Gaussian-shaped barriers at maximum power factor as previously described in Fig. 2. It is evident from Fig. 3c that the highest power factor is observed for small values of decay length ($\xi$~0.8 nm). Perfect square wells ($\xi$=0) perform slightly lower, which demonstrates that sharp edges in the well shape could degrade performance slightly. This degradation effect due to extremely sharp features is interesting, and could be attributed to reduction of the conductivity due to quantum mechanical reflections and oscillations caused by the sharp features as described in Ref. [23], however appears to result in much smaller losses than the other effects discussed here and is thus no explored further in this work. The maximum value is similar to that of the Gaussian-shaped geometry power factor (black lines). As $\xi$ increases the power factor drops significantly, approaching towards the uniform single barrier channel performance (magenta line). The conductance and Seebeck coefficients in Fig. 3a and 3b are also similar to the corresponding values for the best Gaussian profile (black lines) for small decay lengths. The weak variation of $G$ and $S$ with $\xi$ allows the power factor to remain high even up to values $\xi$~2 nm, for which the wells are moderately distorted (middle inset of Fig. 3c). A very large distortion of the wells, up to $\xi$~6 nm (see right inset of Fig. 3c), is required for the power factor $GS^2$ to reduce to the values of the single barrier geometry (magenta-dashed line).

Barrier thickness and the influence of tunneling: An important observation regarding the results described in Fig. 2 for the shape of both the barrier and well, compared to the results in Fig. 3 where only the well is changed, is the much stronger sensitivity of both $G$ and $S$ to the barrier shape compared to the well shape. The influence of tunneling at the top of the barrier (in the results in Fig. 2) can lead to large conductance, but low Seebeck coefficient. In Fig. 4 we emphasize the importance of tunneling by showing the power factor of some of the structures from Fig. 3 versus barrier width, $W$. For smaller barrier thicknesses, below $W=3$ nm, tunneling degrades the Seebeck coefficient strongly. For larger thicknesses, the low energy/velocity carriers on top of the barriers increase the resistance of the overall material. The power factor peaks somewhere around $W=3$nm. Thus, the optimal barrier needs to be thick enough for tunneling to be prevented, but thin enough for its resistance to remain low [23].

Variations in design parameters: To this point, we have shown how deviations of the barrier and well shapes affect the power factor of superlattice nanostructures, and how any advantages compared to the uniform, single barrier thermoelectric operation can be suppressed. Additionally, though, in a real material all design parameters are subject to process variations. The sizes and shapes of the barriers and wells, as well as the barrier heights ($V_B$) can be statistically varying along the length of the material. Variations of some of these parameters could have only a minor effect on the thermoelectric power factor, but variations of others could have a significant influence.

Figures 5a, 5b, and 5c show the conductance, Seebeck coefficient, and power factor of materials in which the parameters described above statistically vary along the transport direction (see right inset of Fig. 5c). The thermoelectric coefficients are plotted versus the degree of statistical variability as a percentage of the initial value. The pristine structure has square shaped wells and barriers with $L_W=20$ nm, $W=3$ nm, $E_F=0.14$ eV, and $V_B=0.16$ eV. The effect of variations in the well width $L_W$ is shown in blue ($\Delta L_W$), the effect of variations in the barrier width $W$ is shown in red ($\Delta W$), the effect of variations in the barrier height $V_B$ in black ($\Delta V_B$), and the overall effect in varying all the above parameters simultaneously, as well as the decay length $\xi$, in green. We simulate structures in which we allow variations up to 30% in the design parameters, and extract statistics from at least 100 geometry realizations for every data point presented in Fig. 5.

The left side of Fig. 5, for zero variation, indicates the performance of the initial, pristine superlattice material, which turns out to be the highest. As the degree of variation increases, the conductance $G$ drops in all cases (Fig. 5a), the Seebeck coefficient $S$ increases (Fig. 5b), but overall the power factor $GS^2$ drops, following the conductance trend. Not all parameters degrade the power factor equally. It turns out that variations in the widths of the wells $L_W$ and barriers $W$ (as long as significant tunneling is not introduced), only affect the power factor weakly. At the maximum variation we simulate, the conductivity drops by ~25%, the Seebeck coefficient increases by 10%, which results in a minor reduction in the power factor (Fig. 5c).

Very large power factor degradation, however, is observed with variation in the barrier height $V_B$. At 10% variation in the barrier height, which corresponds to a variation of 16 meV (less than $k_BT$=26 meV), the advantage of the superlattice geometry is already entirely erased as shown in Fig. 5c (at 10% average variation, $GS^2$ crosses the magenta line for the performance of the single barrier thermoelectric case material). For variations up to 30% (~40 meV, or somewhat less than $2k_BT$), the power factor of the superlattice material drops to even half of the corresponding single barrier material value. Noticeably, the performance reduction due to variations in $V_B$ dominates that of all other parameter variations combined. This also indicates that the influence of variations in the shapes of the wells is insignificant to the power factor, as expected following the results of Fig. 3. In order to quantify our understanding on the effect $V_B$ variations on the power factor, in the left inset of Fig. 5c we plot the power factor of the structures simulated, versus the maximum barrier height in the structure. Interestingly, the power factors follow a descending trend (black dots), indicating that the overall performance is dominated by the highest barrier height in the channel alone. Indeed, the red-dashed line indicates the power factor of the pristine structure, but with the middle barrier alone raised to the value of the highest barrier. This forms an envelope to the results of the structures with varying features, again indicating that the single highest barrier dominates the performance.

## IV. Discussion

The results of Fig. 5c clearly show that well-designed superlattices could result in ~40% thermoelectric power factor improvements compared to materials with a uniform underlying potential. For this, a series of parameters needs to be carefully calibrated as mentioned earlier (i.e. semi-ballistic wells, proper positioning of the Fermi level with respect to the barriers, proper barrier width). In addition, however, for these improvements to be realized, a very good control of the barrier heights needs to be achieved. Large effort is currently being devoted in achieving high power factors in such geometries, but in several occasions, due to variability in material fabrication, perfect material realization according to the optimal specifications cannot be obtained. In this work, we stress the importance in achieving well-controlled barrier heights above all other process parameter variations. We also need to stress that superlattices, and nanocomposites in general, provide high $ZT$ figures of merit as a consequence of their extremely low thermal conductivities [5, 14, 33-34], as well as the non-uniformity of the spatial thermal conductivity [13,25], and these in and of themselves suggest they are indeed very promising thermoelectric materials. Achieving additional power factor benefits through energy filtering, however, seems to require more control over several design parameters and their variability and could be a more difficult task [35-36]. We do not consider the benefits from low and non-uniform spatial thermal conductivities in this work, but it might be the case that the power factor reduction under the influence of parameter variability could then be compensated, and high $ZTs$ could be achieved. Another interesting point is that the introduction of superlattices targeted the improvement of the Seebeck coefficient through energy filtering. However, it seems that when considering the influence of variability, it is the behavior of $G$ which dominates the $GS^2$, rather than that of $S$.

With regards to the constant effective mass used, we want to stress that in this work we limit the parameters of variation we consider to geometrical features and potential profile shapes. One can of course reasonably consider variations in the effective mass, which will also suggest variations in the electro-phonon interaction strengths as well, but these will largely increase the parameter space of possible parameter variation. Nevertheless, random variations in the effective masses locally in random places of the superlattice due to the presence of varying strain fields or imperfect alloying, etc. could exist in real structures. Such non-uniformities bring almost linear (or small) changes to the transport features of the channel, i.e. they change the carrier velocities slightly, they introduce weak scattering centers, and they create a non-smooth potential profile. Thus, we would expect that they will have a qualitatively moderate degrading effect to the power factor, similar to the effect that variations in the barrier shape introduce, as we present above. The qualitatively strongest effect will come from variations of the barrier height, as concluded above. Note that these random variations we describe do not correspond to possible well-controlled variation of the effective mass between barriers and wells. In that case one could find an optimal relation between the masses in the wells and barriers as described in Ref. [37] which would provide higher power factors, but variations in the values of those masses (under zero barrier height variation), would also introduce moderate degradation in performance.

## V. Conclusions

In conclusion, we have investigated the potential for parameter variability and random variation to destroy any gains in the thermoelectric power factor due to energy filtering in cross-plane superlattices composed of nanometer size wells and barriers. We employed the quantum mechanical non-equilibrium Green's function method including electron scattering with acoustic and optical phonons. Starting from an optimized superlattice pristine material geometry which shows ~40% power factor improvement compared to the uniform material, we show that any deviations from the ideal design can significantly minimize or entirely eliminate the gains resulting from the multi-barrier geometry. We showed that variation in the barrier shape and width in a way that it allows for tunneling is especially detrimental to the superlattice power factor. A large degradation to the power factor is also observed upon statistical variations in the barrier heights along the transport path, which needs to be avoided if benefits to the power factor are to be realized. Variations in the width and shape of the wells along the material transport direction, on the other hand, do not affect the power factor significantly.

*Acknowledgement:* Mischa Thesberg, Mahdi Pourfath, and Hans Kosina were supported by the Austrian Science Fund (FWF) contract P25368-N30. The computational results presented have been achieved in part using the Vienna Scientific Cluster (VSC).

Figure 1:

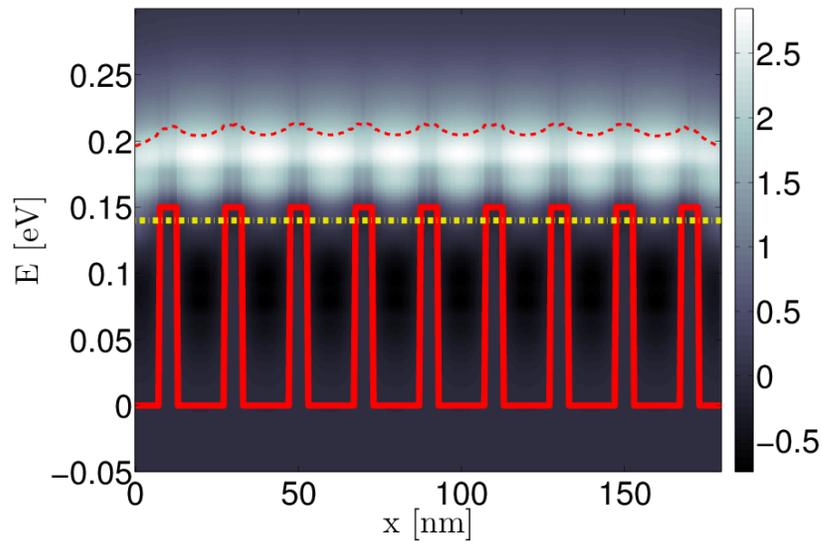

Figure 1 caption:

(a) The band diagram of the superlattice materials under consideration, consisting of a series of potential wells and barriers. The wells have width $L_W$=20 nm and the barriers width $W$=3 nm. The Fermi level is shown by the yellow-dashed line. The colormap shows the current energy spectrum through the superlattice material (the average energy of the current is shown by the red-dashed line). Most of the current passes over the barriers, however, significant energy relaxation is observed in the wells. The material is designed to have 50% ballisticity in electron transport in the wells.

Figure 2:

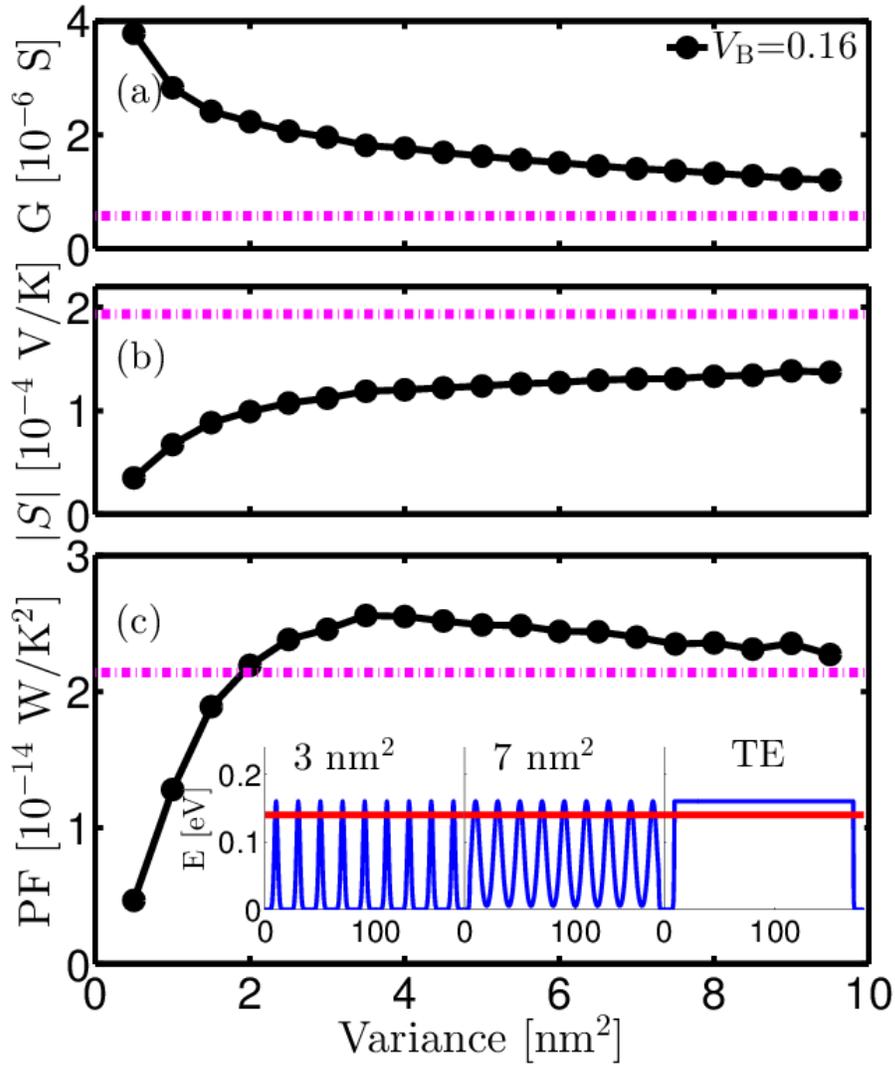

Figure 2 caption:

The influence of deviations of the shape of the barriers and wells of the pristine superlattice material from a square into a Gaussian-like shape. (a) The electrical conductance, (b) the Seebeck coefficient, and (c) the power factor versus the Gaussian profile variance. The barrier height is $V_B$=0.16 eV. The conductance, Seebeck coefficient and power factor of the single barrier channel (usual *bulk thermoelectric* operation) are indicated by the magenta-dashed lines. Insets of (c) from left to right: The potential profiles in channels with Gaussian shaped barriers and wells of variance 3nm$^2$ (δ-function like barriers), variance 7nm$^2$, and the case of a single barrier (bulk thermoelectric operation).

Figure 3:

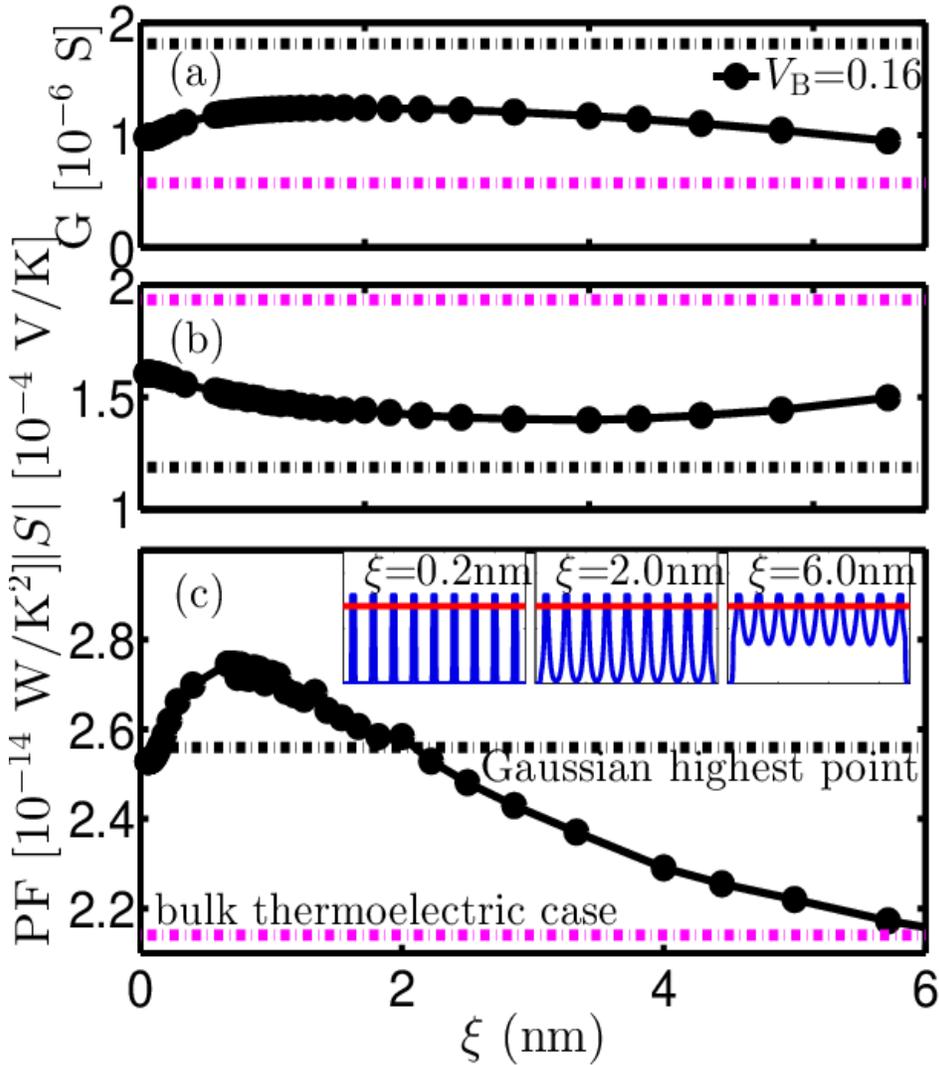

Figure 3 caption:

The influence of deviations of the shape of the well alone of the pristine superlattice material from a square into a curved shape. (a) The electronic conductance, (b) the Seebeck coefficient, and (c) the power factor versus the curved profile decay length $\xi$. The barrier height is $V_B$=0.16 eV. The corresponding quantities of the single barrier uniform channel are indicated by the magenta-dashed lines. The corresponding quantities at the maximum power factor for the channels in Fig. 2 (Gaussian profiles) are indicated by the black-dashed lines. Insets of Fig. 3c from left to right: The potential profile in a channel with weakly curved well potential $\xi$=0.2 nm, distorted $\xi$=2 nm, and heavily distorted $\xi$=6 nm.

Figure 4:

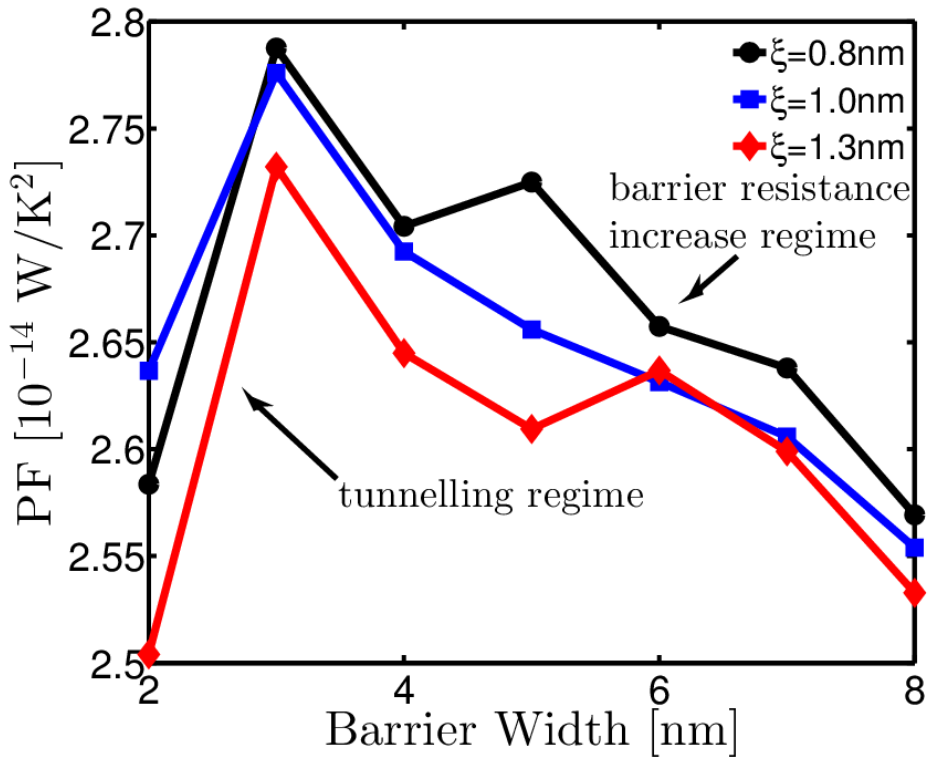

Figure 4 caption:

The thermoelectric power factor of the materials with curve-shaped potential wells versus barrier width for different decay length values. At the left side (for thin barriers) the power factor suffers from tunneling, whereas at the right side (wide barriers) it suffers from increased barrier resistance.

Figure 5:

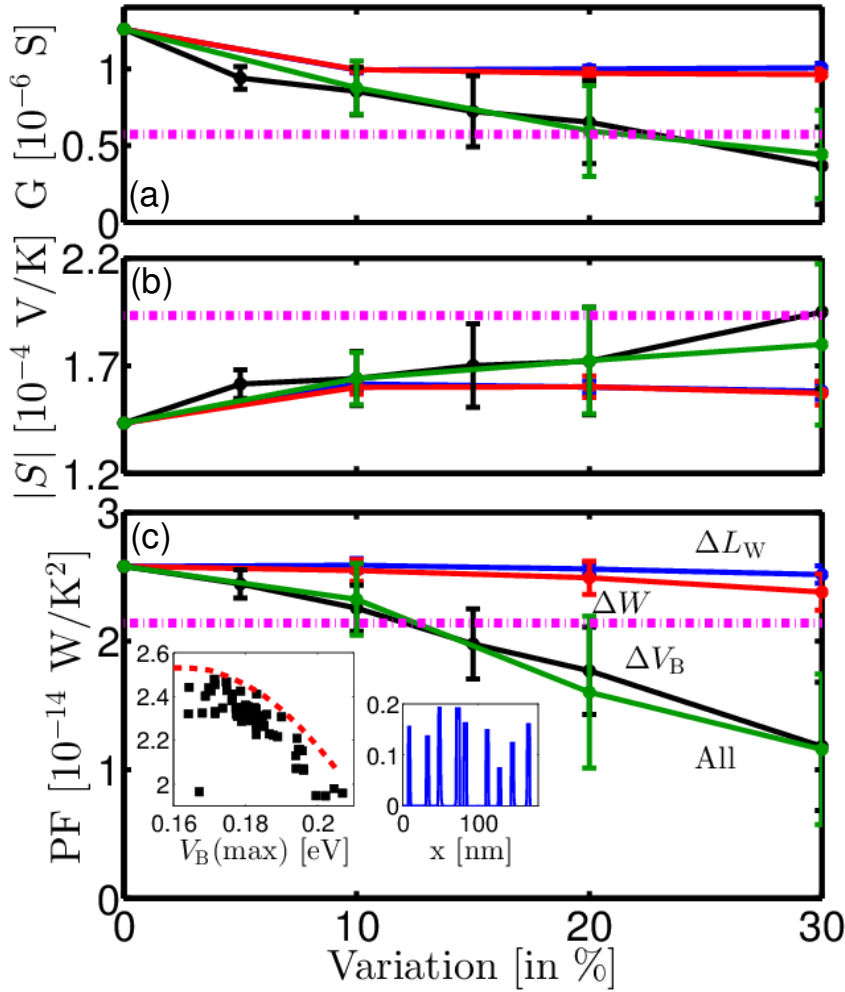

Figure 5 caption:

The influence of the channel imperfections on the thermoelectric coefficients. (a) The electrical conductance, (b) the Seebeck coefficient, and (c) the power factor versus the percentage of the statistical variation from the nominal values. Variations in the width of the wells ($\Delta L_W$-blue lines), the width of the barriers ($\Delta W$-red lines), the barrier height ($\Delta V_B$-black lines), and variations in all parameters combined (green lines) are shown. Statistics for each data point were extracted from simulations of at least 100 randomized channel realizations. Insets of (c): The left shows the power factor of the data in the same units and label as in (c) versus the highest barrier height (red line is the power factor of the pristine superlattice with only the central barrier raised). The right inset shows a sample geometry with 30% variation in all parameters.